\documentclass[manuscript,screen]{acmart}

\AtBeginDocument{%
  \providecommand\BibTeX{{%
    \normalfont B\kern-0.5em{\scshape i\kern-0.25em b}\kern-0.8em\TeX}}}

\setcopyright{acmcopyright}
\copyrightyear{2021}
\acmYear{2021}
\acmDOI{10.1145/1122445.1122456}

\acmConference[SimuRec '21]{SimuRec '21: The SimuRec workshop held in conjunction with the 15th ACM Conference on Recommender Systems (RecSys), 2021, in Amsterdam, Netherlands.}{Sep 27-- Oct 01, 2021}{Amsterdam, Netherlands (online)}
\acmBooktitle{SimuRec '21: The SimuRec workshop held in conjunction with the 15th ACM Conference on Recommender Systems (RecSys),
  Sep 27-- Oct 01, 2021, Amsterdam, Netherlands (online)}

\usepackage[ruled]{algorithm2e} 
\usepackage[noend]{algpseudocode}

\begin{document}

\title[Simulations for novel problems in recommendation]{Simulations for novel problems in recommendation: analyzing misinformation and data characteristics}

\author{Alejandro Bellog\'in}
\authornote{Both authors contributed equally to this research.}
\email{alejandro.bellogin@uam.es}
\orcid{0000-0001-6368-2510}
\affiliation{%
  \institution{Universidad Aut\'onoma de Madrid}
  \country{Spain}
}
\author{Yashar Deldjoo}
\authornotemark[1]
\email{yashar.deldjoo@poliba.it}
\orcid{0000-0002-6767-358X}
\affiliation{%
 \institution{Politecnico di Bari}
 \country{Italy}
}

\begin{abstract}
In this position paper, we discuss recent applications of simulation approaches for recommender systems tasks.
In particular, we describe how they were used to analyze the problem of misinformation spreading and understand which data characteristics affect the performance of recommendation algorithms more significantly.
We also present potential lines of future work where simulation methods could advance the work in the recommendation community.
\end{abstract}



\maketitle

\section{Importance of sampling in recommendation}
\label{sec:intro}
Recent years have witnessed an explosion where simulations have been used in several aspects of Recommender Systems (RS), either inspired from Machine Learning (ML) or Information Retrieval (IR) problems -- where simulations have been used for a long time, as in click models \cite{DBLP:series/synthesis/2015Chuklin} or learning to rank \cite{DBLP:journals/tois/AiYWM21} approaches --, or to attack inherent and concrete issues that are prevalent in RS, such as data scarcity and sparsity.
%

In general, sampling in recommendation has been used to open up (or \textit{simulate}) evaluation scenarios.
For example, in \cite{DBLP:conf/recsys/KluverK14} the authors used it to downsample the observed interactions and generate different scenarios of cold-start.
We also found works where sampling is employed to analyze and understand the evaluation procedure, as in \cite{DBLP:conf/recsys/ValcarceBPC18} where it is used to perform a robustness analysis, in \cite{DBLP:conf/recsys/AnelliNSPR19} where the authors extend the previous work to study hyperparameter optimization, or
to directly debias the evaluation, as described in \cite{DBLP:conf/sac/CarraroB20}.

However, there is also a growing body of literature that has been using sampling for learning preferences, as in the well-known Bayesian Personalized Ranking (BPR) algorithm \cite{DBLP:conf/uai/RendleFGS09}, which uses sampling to select pairs of items which are then provided to the algorithm.
However, it is not obvious how this sampling should be done so that it always work, and several approaches have been investigated which differ in effectiveness and biases -- such as popularity -- learned by the method \cite{jadidinejad2019sensitive,10.1145/3450289}.
Recent works that also rely on samplings or simulations are those where reinforcement learning is applied to recommendation \cite{DBLP:journals/ftir/Glowacka19} or to its evaluation \cite{DBLP:conf/recsys/HuangORH20}.
Similarly, whenever the IR or ML approaches mentioned before have even been applied to recommendation, as in \cite{DBLP:conf/ecir/HofmannSBR14}, the sampling process is key for a successful translation of the approach from one area to the other. 

In the rest of this paper, we present two use cases where the authors have recently applied data sampling on recommendation tasks with different goals in each case.
We later discuss the main lessons learned in this process and the advantages of using simulations or sampling strategies.
We end the paper with several open questions or ideas to explore in the future.

\section{Successful applications of sampling}
\label{sec:app}
In this section, we describe two applications of data sampling on recent works performed by the authors.
Section \ref{ss:misinfo} shows how sampling allowed to analyze the effect of misinformation in recommender systems, whereas Section \ref{ss:char} describes a sampling strategy that is used to infer the impact of data characteristics in recommendation performance.

\subsection{Analyzing misinformation}
\label{ss:misinfo}
In \cite{DBLP:journals/corr/abs-2103-14748}, we analyzed the effect that some recommendation algorithms cause on the amplification of misinformation.
For this, we first created a dataset by merging information from fact-checkers and Twitter.
Then, to simulate a wide range of situations -- not only the one captured at the moment of collecting the data -- we used Algorithm~\ref{alg:ratio} to generate user profiles with a pre-defined set of constraints, in particular, the proportion of users who shared some misinformative items. In this way, we would simulate a population where 20\% (or 50\% or 80\%) of users share this type of item.
This also gives us more control on the amount of information received by the algorithms in terms of sparsity or other important dataset statistics.

\newlength{\commentWidth}
\setlength{\commentWidth}{3.6cm}
\newcommand{\atcp}[1]{\tcp*[r]{\makebox[\commentWidth]{#1\hfill}}}

\begin{algorithm}[t]
\caption{Ratio-based user profile generator}
\label{alg:ratio}
\SetKwProg{generate}{Function \emph{generate}}{}{end}

\generate{user u, ratio r}{
     neg $\gets$ \{ i $\in$ u: i \text{ is misinformative } \} \atcp{Negative claims}
     neu $\gets$ u $\setminus$ neg \atcp{Neutral claims} 
     desNeg $\gets$ r $\cdot$ $|$u$|$ \atcp{Desired negative ratio}
     desNeu $\gets$ (1 - r) $\cdot$ $|$u$|$ \atcp{Desired neutral ratio}
     \While{ (desNeg $>$ $|$neg$|$) OR (desNeu $>$ $|$neu$|$)}{
     \If{ desNeg $>$ $|$neg$|$ \atcp{Downsampling negative} }{ 
          desNeg $\gets$ desNeg - 1\;
      }
     \If{ desNeu $>$ $|$neu$|$ \atcp{Downsampling neutral} }{
          desNeu $\gets$ desNeu - 1\;
      }
      newTotal $\gets$ desNeg + desNeu\;
      desNeg $\gets$ r $\cdot$ newTotal\;
      desNeu $\gets$ (1 - r) $\cdot$ newTotal\;
     }
     userProfile(u) $\gets$ sample(neg, desNeg) $\cup$ sample(neu, desNeu)\;
}
\end{algorithm}

It is interesting to note that simulation has been very recently proposed in \cite{DBLP:journals/corr/abs-2107-08959} also to study the societal impact of recommender systems. By a general event-driven simulation model, the authors analyze some case studies and discuss the implications for reproducibility such a framework may have.

\subsection{Understanding data characteristics}
\label{ss:char}
In \cite{ipm2021}, and as an extension of \cite{DBLP:conf/sigir/DeldjooNSM20}, we aim to better understand the impact of an array of characteristics in a dataset with respect to accuracy and fairness (in \cite{ipm2021}) or robustness (in \cite{DBLP:conf/sigir/DeldjooNSM20}).
With this goal in mind, we develop an explanatory framework using regression models in its core, a methodology originally proposed in \cite{DBLP:journals/tmis/AdomaviciusZ12}.
In this framework, we test whether a set of data characteristics (independent variables of the regression model) are related to a given performance metric (the dependent variable).
To obtain enough points so that the framework could produce significant results, we need to \textit{simulate} $N$ different datasets ($N=600$ in both our papers). These simulated datasets are generated from sampling the original dataset as described in Algorithm \ref{alg:sampling}, which produces smaller datasets with slightly different characteristics in each case, although satisfying some constraints that allow the dataset to be useful in the analysis -- for example, by restricting the maximum number of items (via $\tau_i$) on these smaller datasets. It is possible to perform these experiments by considering different types of item content \cite{deldjoo2021content,DBLP:conf/esws/AnelliDNSM20,deldjoo2018content} along different evaluation objective, e.g., accuracy, robustness \cite{deldjoo2019assessing,DBLP:conf/esws/AnelliDNSM20,DBLP:conf/sigir/DeldjooNSM20}, privacy \cite{anelli2021pursuing,slokom2020partially}, or fairness \cite{ipm2021,lesota2021analyzing}.

\begin{algorithm}[t]
\caption{Sample generation procedure}
\label{alg:sampling}
\SetKwProg{sampling}{Function \emph{data-sampling}}{}{end}

\sampling{URM user-item rating matrix}{
        $n_{u} \gets$ number of users of the URM\; 
        $n_{i} \gets$ number of users of the URM\;
        $n_{r} \gets$ number of ratings of the URM\;
        $\tau_{u} \gets$ constraint on average number of ratings for users\;
        $\tau_{i} \gets$ constraint on maximum number of items\;
        $ n \gets 1$ \;
        \While{ $n \leq N$ }{
            Random shuffle the rows of the URM\;
            $n_{u} \gets rnd([100, n_u])$\;
            $n_{i} \gets rnd([100, n_i])$\;
            $urm_{n} \gets$ Selection of $n_{u}$, $n_{i}$ from shuffled URM\;
            \If{ $\frac{n_{r}}{n_{u}}< \tau_u$ or $n_{i} > \tau_i$ }{
                $ n \gets n + 1$\;
            }
          }
        \textbf{Output:} $N$ sub-datasets ($urm_{n}$)\;
}
\end{algorithm}

\section{Discussion}
\label{sec:disc}
From the applications presented in Section \ref{sec:app} we have learned that simulation, and in particular, data sampling, is complex, even though it might be beneficial -- sometimes the only tool to produce the data points needed for an analysis.
Its difficulty relies on sampling the data in a significant but fair and realistic way, without introducing new biases, as acknowledged in recent works such as \cite{jadidinejad2019sensitive}.
For example, in the case described in Section\ref{ss:char} we experimented with imposing additional constraints to obtain simulated datasets with a particular density, number of ratings per user, or number of items.

Based on our experience, we foresee a continuous use of simulation techniques, such as data sampling or synthetic data, to experiment with novel evaluation conditions, such as those described before.
In particular, consider the limitations of public datasets, with a fixed number of attributes for users and items and a (sometimes) a small number of interactions, these strategies will allow evaluating new conditions with little effort -- or at least, with less effort than that of creating datasets including all the required information.

\section{Open questions}
\label{sec:future}
Throughout this position paper, we have presented several situations where simulations or sampling strategies have been successfully applied to recommendation. However, we believe there is room for improvement, and several aspects remain unexplored, both in our works and in the community, under a more general perspective.
%
First, regarding how to extend our sampling approaches, it would be more realistic if the temporal dimension is incorporated in the process, so that the interactions follow a \textit{compatible} order with the one in the original dataset.
When done properly, this would allow generating and studying feedback loops, like those created by reinforcement learning algorithms, but at a higher, more global level.
As observed recently, and in agreement with our misinformation analysis \cite{DBLP:journals/corr/abs-2103-14748}, some algorithms are more prone to reproduce biases at each feedback loop \cite{DBLP:conf/cikm/MansouryAPMB20}.

Regarding our second application (see Section \ref{ss:char}), adding more variability and flexibility in the types of datasets being generated would allow us to address more complex questions.
In particular, we envision a definition of user types (or \textit{personas}) which are then simulated or sampled at different rates, either randomly or controlled via some parameter.
Similarly, generating content features realistically would help go beyond collaborative filtering algorithms and test content-based methods at varying levels of information, quality, and sparsity \cite{deldjoo2021content}. For this, recent advances from the Natural Language Processing community could be convenient, where Neural Networks may generate realistic pieces of text in several domains \cite{DBLP:conf/naacl/DevlinCLT19,DBLP:conf/nips/BrownMRSKDNSSAA20}.
Besides content attributes, including sensitive attributes in the set of controled (or simulated) information to be generated will allow to explore fairness \cite{deldjoo2021flexible} analyses in a more comprehensive way than what is being done right now.
Finally, an interesting perspective that could be promising is to shift the focus of the application, as mentioned earlier, from the data to the algorithms so that the sampling strategy instead of sampling data should sample algorithms. In this way, the input data would be fixed, and the data points for the regression analysis would be obtained from a wide selection of algorithms, probably with different hyperparameters, to increase their variability. Such an approach would have connections with automatic hyperparameter tuning techniques like Bayes Optimization \cite{DBLP:conf/nips/BergstraBBK11}, but it would be applied to solve a different problem and in a more extensive search space, as the type of algorithm will also be part of the sampled variables \cite{DBLP:journals/eor/BengioLP21}.

From a more general perspective, we believe several open questions need to be considered when doing simulations or, in particular, some data sampling.
An important aspect that should be considered is that of reproducibility \cite{DBLP:journals/tois/DacremaBCJ21}. While usually sampling -- or simulations in general -- are random in nature, this hinders reproducing other people's works. However, by sharing the code -- where customizable seeds are included wherever is needed -- and/or the generated simulated/sampled datasets or the scripts used, the potential to reproduce these types of works should increase \cite{DBLP:journals/corr/abs-2102-00482}.
By addressing the reproducibility problem, a related issue we have also striven to present properly in the past could be solved. Here we refer to present all the technical decisions involved in a sampling strategy carefully and as detailed as possible. Sometimes pseudocodes or algorithms do not have the granularity level needed to specify implementation details that may greatly impact how the sampling or the simulation is generated.

At the same time, although there are some works already addressing this issue (see \cite{DBLP:conf/recsys/HuangORH20}), we believe it is very important to understand the potential biases that a particular simulation may be introducing in the generated data. This may be intentional but, usually, there should be some mechanism to avoid them.
Related to this problem, there should exist a definition of when a simulation can be considered faithful to the original (or intended) data, which we have referred throughout this work as \textit{realistic}.
Without such definition, we might end up with data samples that are too different (or different in specific, key aspects) from the actual data, hence not satisfying their underlying assumptions or constraints.
This also applies when generating synthetic data, a task not so popular in recommendation because of its difficulty, but where some efforts have been reported in the last years \cite{DBLP:conf/recsys/Slokom18}.

\bibliographystyle{ACM-Reference-Format}
\bibliography{sample-base}


\begin{thebibliography}{33}


\ifx \showCODEN    \undefined \def \showCODEN     #1{\unskip}     \fi
\ifx \showDOI      \undefined \def \showDOI       #1{#1}\fi
\ifx \showISBNx    \undefined \def \showISBNx     #1{\unskip}     \fi
\ifx \showISBNxiii \undefined \def \showISBNxiii  #1{\unskip}     \fi
\ifx \showISSN     \undefined \def \showISSN      #1{\unskip}     \fi
\ifx \showLCCN     \undefined \def \showLCCN      #1{\unskip}     \fi
\ifx \shownote     \undefined \def \shownote      #1{#1}          \fi
\ifx \showarticletitle \undefined \def \showarticletitle #1{#1}   \fi
\ifx \showURL      \undefined \def \showURL       {\relax}        \fi
\providecommand\bibfield[2]{#2}
\providecommand\bibinfo[2]{#2}
\providecommand\natexlab[1]{#1}
\providecommand\showeprint[2][]{arXiv:#2}

\bibitem[\protect\citeauthoryear{Adomavicius and Zhang}{Adomavicius and
  Zhang}{2012}]%
        {DBLP:journals/tmis/AdomaviciusZ12}
\bibfield{author}{\bibinfo{person}{Gediminas Adomavicius} {and}
  \bibinfo{person}{Jingjing Zhang}.} \bibinfo{year}{2012}\natexlab{}.
\newblock \showarticletitle{Impact of data characteristics on recommender
  systems performance}.
\newblock \bibinfo{journal}{\emph{{ACM} Trans. Manag. Inf. Syst.}}
  \bibinfo{volume}{3}, \bibinfo{number}{1} (\bibinfo{year}{2012}),
  \bibinfo{pages}{3:1--3:17}.
\newblock
\urldef\tempurl%
\url{https://doi.org/10.1145/2151163.2151166}
\showDOI{\tempurl}


\bibitem[\protect\citeauthoryear{Ai, Yang, Wang, and Mao}{Ai
  et~al\mbox{.}}{2021}]%
        {DBLP:journals/tois/AiYWM21}
\bibfield{author}{\bibinfo{person}{Qingyao Ai}, \bibinfo{person}{Tao Yang},
  \bibinfo{person}{Huazheng Wang}, {and} \bibinfo{person}{Jiaxin Mao}.}
  \bibinfo{year}{2021}\natexlab{}.
\newblock \showarticletitle{Unbiased Learning to Rank: Online or Offline?}
\newblock \bibinfo{journal}{\emph{{ACM} Trans. Inf. Syst.}}
  \bibinfo{volume}{39}, \bibinfo{number}{2} (\bibinfo{year}{2021}),
  \bibinfo{pages}{21:1--21:29}.
\newblock
\urldef\tempurl%
\url{https://doi.org/10.1145/3439861}
\showDOI{\tempurl}


\bibitem[\protect\citeauthoryear{Anelli, Belli, Deldjoo, Di~Noia, Ferrara,
  Narducci, and Pomo}{Anelli et~al\mbox{.}}{2021}]%
        {anelli2021pursuing}
\bibfield{author}{\bibinfo{person}{Vito~Walter Anelli}, \bibinfo{person}{Luca
  Belli}, \bibinfo{person}{Yashar Deldjoo}, \bibinfo{person}{Tommaso Di~Noia},
  \bibinfo{person}{Antonio Ferrara}, \bibinfo{person}{Fedelucio Narducci},
  {and} \bibinfo{person}{Claudio Pomo}.} \bibinfo{year}{2021}\natexlab{}.
\newblock \showarticletitle{Pursuing Privacy in Recommender Systems: the View
  of Users and Researchers from Regulations to Applications}. In
  \bibinfo{booktitle}{\emph{Fifteenth ACM Conference on Recommender Systems}}.
  \bibinfo{pages}{838--841}.
\newblock


\bibitem[\protect\citeauthoryear{Anelli, Deldjoo, Noia, Sciascio, and
  Merra}{Anelli et~al\mbox{.}}{2020}]%
        {DBLP:conf/esws/AnelliDNSM20}
\bibfield{author}{\bibinfo{person}{Vito~Walter Anelli}, \bibinfo{person}{Yashar
  Deldjoo}, \bibinfo{person}{Tommaso~Di Noia}, \bibinfo{person}{Eugenio~Di
  Sciascio}, {and} \bibinfo{person}{Felice~Antonio Merra}.}
  \bibinfo{year}{2020}\natexlab{}.
\newblock \showarticletitle{SAShA: Semantic-Aware Shilling Attacks on
  Recommender Systems Exploiting Knowledge Graphs}. In
  \bibinfo{booktitle}{\emph{The Semantic Web - 17th International Conference,
  {ESWC} 2020, Heraklion, Crete, Greece, May 31-June 4, 2020, Proceedings}}
  \emph{(\bibinfo{series}{Lecture Notes in Computer Science},
  Vol.~\bibinfo{volume}{12123})}. \bibinfo{publisher}{Springer},
  \bibinfo{pages}{307--323}.
\newblock
\urldef\tempurl%
\url{https://doi.org/10.1007/978-3-030-49461-2\_18}
\showDOI{\tempurl}


\bibitem[\protect\citeauthoryear{Anelli, Noia, Sciascio, Pomo, and
  Ragone}{Anelli et~al\mbox{.}}{2019}]%
        {DBLP:conf/recsys/AnelliNSPR19}
\bibfield{author}{\bibinfo{person}{Vito~Walter Anelli},
  \bibinfo{person}{Tommaso~Di Noia}, \bibinfo{person}{Eugenio~Di Sciascio},
  \bibinfo{person}{Claudio Pomo}, {and} \bibinfo{person}{Azzurra Ragone}.}
  \bibinfo{year}{2019}\natexlab{}.
\newblock \showarticletitle{On the discriminative power of hyper-parameters in
  cross-validation and how to choose them}. In
  \bibinfo{booktitle}{\emph{Proceedings of the 13th {ACM} Conference on
  Recommender Systems, RecSys 2019, Copenhagen, Denmark, September 16-20,
  2019}}, \bibfield{editor}{\bibinfo{person}{Toine Bogers},
  \bibinfo{person}{Alan Said}, \bibinfo{person}{Peter Brusilovsky}, {and}
  \bibinfo{person}{Domonkos Tikk}} (Eds.). \bibinfo{publisher}{{ACM}},
  \bibinfo{pages}{447--451}.
\newblock
\urldef\tempurl%
\url{https://doi.org/10.1145/3298689.3347010}
\showDOI{\tempurl}


\bibitem[\protect\citeauthoryear{Bellog{\'{\i}}n and Said}{Bellog{\'{\i}}n and
  Said}{2021}]%
        {DBLP:journals/corr/abs-2102-00482}
\bibfield{author}{\bibinfo{person}{Alejandro Bellog{\'{\i}}n} {and}
  \bibinfo{person}{Alan Said}.} \bibinfo{year}{2021}\natexlab{}.
\newblock \showarticletitle{Improving Accountability in Recommender Systems
  Research Through Reproducibility}.
\newblock \bibinfo{journal}{\emph{CoRR}}  \bibinfo{volume}{abs/2102.00482}
  (\bibinfo{year}{2021}).
\newblock
\showeprint[arxiv]{2102.00482}
\urldef\tempurl%
\url{https://arxiv.org/abs/2102.00482}
\showURL{%
\tempurl}


\bibitem[\protect\citeauthoryear{Bengio, Lodi, and Prouvost}{Bengio
  et~al\mbox{.}}{2021}]%
        {DBLP:journals/eor/BengioLP21}
\bibfield{author}{\bibinfo{person}{Yoshua Bengio}, \bibinfo{person}{Andrea
  Lodi}, {and} \bibinfo{person}{Antoine Prouvost}.}
  \bibinfo{year}{2021}\natexlab{}.
\newblock \showarticletitle{Machine learning for combinatorial optimization:
  {A} methodological tour d'horizon}.
\newblock \bibinfo{journal}{\emph{Eur. J. Oper. Res.}} \bibinfo{volume}{290},
  \bibinfo{number}{2} (\bibinfo{year}{2021}), \bibinfo{pages}{405--421}.
\newblock
\urldef\tempurl%
\url{https://doi.org/10.1016/j.ejor.2020.07.063}
\showDOI{\tempurl}


\bibitem[\protect\citeauthoryear{Bergstra, Bardenet, Bengio, and
  K{\'{e}}gl}{Bergstra et~al\mbox{.}}{2011}]%
        {DBLP:conf/nips/BergstraBBK11}
\bibfield{author}{\bibinfo{person}{James Bergstra}, \bibinfo{person}{R{\'{e}}mi
  Bardenet}, \bibinfo{person}{Yoshua Bengio}, {and}
  \bibinfo{person}{Bal{\'{a}}zs K{\'{e}}gl}.} \bibinfo{year}{2011}\natexlab{}.
\newblock \showarticletitle{Algorithms for Hyper-Parameter Optimization}. In
  \bibinfo{booktitle}{\emph{Advances in Neural Information Processing Systems
  24: 25th Annual Conference on Neural Information Processing Systems 2011.
  Proceedings of a meeting held 12-14 December 2011, Granada, Spain}},
  \bibfield{editor}{\bibinfo{person}{John Shawe{-}Taylor},
  \bibinfo{person}{Richard~S. Zemel}, \bibinfo{person}{Peter~L. Bartlett},
  \bibinfo{person}{Fernando C.~N. Pereira}, {and} \bibinfo{person}{Kilian~Q.
  Weinberger}} (Eds.). \bibinfo{pages}{2546--2554}.
\newblock
\urldef\tempurl%
\url{https://proceedings.neurips.cc/paper/2011/hash/86e8f7ab32cfd12577bc2619bc635690-Abstract.html}
\showURL{%
\tempurl}


\bibitem[\protect\citeauthoryear{Brown, Mann, Ryder, Subbiah, Kaplan, Dhariwal,
  Neelakantan, Shyam, Sastry, Askell, Agarwal, Herbert{-}Voss, Krueger,
  Henighan, Child, Ramesh, Ziegler, Wu, Winter, Hesse, Chen, Sigler, Litwin,
  Gray, Chess, Clark, Berner, McCandlish, Radford, Sutskever, and Amodei}{Brown
  et~al\mbox{.}}{2020}]%
        {DBLP:conf/nips/BrownMRSKDNSSAA20}
\bibfield{author}{\bibinfo{person}{Tom~B. Brown}, \bibinfo{person}{Benjamin
  Mann}, \bibinfo{person}{Nick Ryder}, \bibinfo{person}{Melanie Subbiah},
  \bibinfo{person}{Jared Kaplan}, \bibinfo{person}{Prafulla Dhariwal},
  \bibinfo{person}{Arvind Neelakantan}, \bibinfo{person}{Pranav Shyam},
  \bibinfo{person}{Girish Sastry}, \bibinfo{person}{Amanda Askell},
  \bibinfo{person}{Sandhini Agarwal}, \bibinfo{person}{Ariel Herbert{-}Voss},
  \bibinfo{person}{Gretchen Krueger}, \bibinfo{person}{Tom Henighan},
  \bibinfo{person}{Rewon Child}, \bibinfo{person}{Aditya Ramesh},
  \bibinfo{person}{Daniel~M. Ziegler}, \bibinfo{person}{Jeffrey Wu},
  \bibinfo{person}{Clemens Winter}, \bibinfo{person}{Christopher Hesse},
  \bibinfo{person}{Mark Chen}, \bibinfo{person}{Eric Sigler},
  \bibinfo{person}{Mateusz Litwin}, \bibinfo{person}{Scott Gray},
  \bibinfo{person}{Benjamin Chess}, \bibinfo{person}{Jack Clark},
  \bibinfo{person}{Christopher Berner}, \bibinfo{person}{Sam McCandlish},
  \bibinfo{person}{Alec Radford}, \bibinfo{person}{Ilya Sutskever}, {and}
  \bibinfo{person}{Dario Amodei}.} \bibinfo{year}{2020}\natexlab{}.
\newblock \showarticletitle{Language Models are Few-Shot Learners}. In
  \bibinfo{booktitle}{\emph{Advances in Neural Information Processing Systems
  33: Annual Conference on Neural Information Processing Systems 2020, NeurIPS
  2020, December 6-12, 2020, virtual}}, \bibfield{editor}{\bibinfo{person}{Hugo
  Larochelle}, \bibinfo{person}{Marc'Aurelio Ranzato}, \bibinfo{person}{Raia
  Hadsell}, \bibinfo{person}{Maria{-}Florina Balcan}, {and}
  \bibinfo{person}{Hsuan{-}Tien Lin}} (Eds.).
\newblock
\urldef\tempurl%
\url{https://proceedings.neurips.cc/paper/2020/hash/1457c0d6bfcb4967418bfb8ac142f64a-Abstract.html}
\showURL{%
\tempurl}


\bibitem[\protect\citeauthoryear{Carraro and Bridge}{Carraro and
  Bridge}{2020}]%
        {DBLP:conf/sac/CarraroB20}
\bibfield{author}{\bibinfo{person}{Diego Carraro} {and} \bibinfo{person}{Derek
  Bridge}.} \bibinfo{year}{2020}\natexlab{}.
\newblock \showarticletitle{Debiased offline evaluation of recommender systems:
  a weighted-sampling approach}. In \bibinfo{booktitle}{\emph{{SAC} '20: The
  35th {ACM/SIGAPP} Symposium on Applied Computing, online event, [Brno, Czech
  Republic], March 30 - April 3, 2020}},
  \bibfield{editor}{\bibinfo{person}{Chih{-}Cheng Hung},
  \bibinfo{person}{Tom{\'{a}}s Cern{\'{y}}}, \bibinfo{person}{Dongwan Shin},
  {and} \bibinfo{person}{Alessio Bechini}} (Eds.). \bibinfo{publisher}{{ACM}},
  \bibinfo{pages}{1435--1442}.
\newblock
\urldef\tempurl%
\url{https://doi.org/10.1145/3341105.3375759}
\showDOI{\tempurl}


\bibitem[\protect\citeauthoryear{Chen, Jiang, Wang, Zhou, Feng, Chen, Ester,
  and He}{Chen et~al\mbox{.}}{2021}]%
        {10.1145/3450289}
\bibfield{author}{\bibinfo{person}{Jiawei Chen}, \bibinfo{person}{Chengquan
  Jiang}, \bibinfo{person}{Can Wang}, \bibinfo{person}{Sheng Zhou},
  \bibinfo{person}{Yan Feng}, \bibinfo{person}{Chun Chen},
  \bibinfo{person}{Martin Ester}, {and} \bibinfo{person}{Xiangnan He}.}
  \bibinfo{year}{2021}\natexlab{}.
\newblock \showarticletitle{CoSam: An Efficient Collaborative Adaptive Sampler
  for Recommendation}.
\newblock \bibinfo{journal}{\emph{ACM Trans. Inf. Syst.}} \bibinfo{volume}{39},
  \bibinfo{number}{3}, Article \bibinfo{articleno}{34} (\bibinfo{date}{May}
  \bibinfo{year}{2021}), \bibinfo{numpages}{24}~pages.
\newblock
\showISSN{1046-8188}
\urldef\tempurl%
\url{https://doi.org/10.1145/3450289}
\showDOI{\tempurl}


\bibitem[\protect\citeauthoryear{Chuklin, Markov, and de~Rijke}{Chuklin
  et~al\mbox{.}}{2015}]%
        {DBLP:series/synthesis/2015Chuklin}
\bibfield{author}{\bibinfo{person}{Aleksandr Chuklin}, \bibinfo{person}{Ilya
  Markov}, {and} \bibinfo{person}{Maarten de Rijke}.}
  \bibinfo{year}{2015}\natexlab{}.
\newblock \bibinfo{booktitle}{\emph{Click Models for Web Search}}.
\newblock \bibinfo{publisher}{Morgan {\&} Claypool Publishers}.
\newblock
\urldef\tempurl%
\url{https://doi.org/10.2200/S00654ED1V01Y201507ICR043}
\showDOI{\tempurl}


\bibitem[\protect\citeauthoryear{Dacrema, Boglio, Cremonesi, and
  Jannach}{Dacrema et~al\mbox{.}}{2021}]%
        {DBLP:journals/tois/DacremaBCJ21}
\bibfield{author}{\bibinfo{person}{Maurizio~Ferrari Dacrema},
  \bibinfo{person}{Simone Boglio}, \bibinfo{person}{Paolo Cremonesi}, {and}
  \bibinfo{person}{Dietmar Jannach}.} \bibinfo{year}{2021}\natexlab{}.
\newblock \showarticletitle{A Troubling Analysis of Reproducibility and
  Progress in Recommender Systems Research}.
\newblock \bibinfo{journal}{\emph{{ACM} Trans. Inf. Syst.}}
  \bibinfo{volume}{39}, \bibinfo{number}{2} (\bibinfo{year}{2021}),
  \bibinfo{pages}{20:1--20:49}.
\newblock
\urldef\tempurl%
\url{https://doi.org/10.1145/3434185}
\showDOI{\tempurl}


\bibitem[\protect\citeauthoryear{Deldjoo, Anelli, Zamani, Bellogin, and
  Di~Noia}{Deldjoo et~al\mbox{.}}{2021a}]%
        {deldjoo2021flexible}
\bibfield{author}{\bibinfo{person}{Yashar Deldjoo},
  \bibinfo{person}{Vito~Walter Anelli}, \bibinfo{person}{Hamed Zamani},
  \bibinfo{person}{Alejandro Bellogin}, {and} \bibinfo{person}{Tommaso
  Di~Noia}.} \bibinfo{year}{2021}\natexlab{a}.
\newblock \showarticletitle{A flexible framework for evaluating user and item
  fairness in recommender systems}.
\newblock \bibinfo{journal}{\emph{User Modeling and User-Adapted Interaction}}
  (\bibinfo{year}{2021}), \bibinfo{pages}{1--47}.
\newblock


\bibitem[\protect\citeauthoryear{Deldjoo, Bellog{\'{\i}}n, and Noia}{Deldjoo
  et~al\mbox{.}}{2021b}]%
        {ipm2021}
\bibfield{author}{\bibinfo{person}{Yashar Deldjoo}, \bibinfo{person}{Alejandro
  Bellog{\'{\i}}n}, {and} \bibinfo{person}{Tommaso~Di Noia}.}
  \bibinfo{year}{2021}\natexlab{b}.
\newblock \showarticletitle{Explaining Recommender Systems Fairness and
  Accuracy through the Lens of Data Characteristics}.
\newblock \bibinfo{journal}{\emph{Inf. Process. Manag.}} \bibinfo{volume}{58},
  \bibinfo{number}{5} (\bibinfo{year}{2021}).
\newblock
\urldef\tempurl%
\url{https://doi.org/10.1016/j.ipm.2021.102662}
\showDOI{\tempurl}


\bibitem[\protect\citeauthoryear{Deldjoo, Di~Noia, and Merra}{Deldjoo
  et~al\mbox{.}}{2019}]%
        {deldjoo2019assessing}
\bibfield{author}{\bibinfo{person}{Yashar Deldjoo}, \bibinfo{person}{Tommaso
  Di~Noia}, {and} \bibinfo{person}{Felice~Antonio Merra}.}
  \bibinfo{year}{2019}\natexlab{}.
\newblock \showarticletitle{Assessing the Impact of a User-Item Collaborative
  Attack on Class of Users}. In \bibinfo{booktitle}{\emph{ImpactRS@RecSys'19
  Workshop on the Impact of Recommender Systems}}.
\newblock


\bibitem[\protect\citeauthoryear{Deldjoo, Noia, Sciascio, and Merra}{Deldjoo
  et~al\mbox{.}}{2020}]%
        {DBLP:conf/sigir/DeldjooNSM20}
\bibfield{author}{\bibinfo{person}{Yashar Deldjoo}, \bibinfo{person}{Tommaso~Di
  Noia}, \bibinfo{person}{Eugenio~Di Sciascio}, {and}
  \bibinfo{person}{Felice~Antonio Merra}.} \bibinfo{year}{2020}\natexlab{}.
\newblock \showarticletitle{How Dataset Characteristics Affect the Robustness
  of Collaborative Recommendation Models}. In
  \bibinfo{booktitle}{\emph{Proceedings of the 43rd International {ACM} {SIGIR}
  conference on research and development in Information Retrieval, {SIGIR}
  2020, Virtual Event, China, July 25-30, 2020}}. \bibinfo{publisher}{{ACM}},
  \bibinfo{pages}{951--960}.
\newblock
\urldef\tempurl%
\url{https://doi.org/10.1145/3397271.3401046}
\showDOI{\tempurl}


\bibitem[\protect\citeauthoryear{Deldjoo, Schedl, Cremonesi, and Pasi}{Deldjoo
  et~al\mbox{.}}{2018}]%
        {deldjoo2018content}
\bibfield{author}{\bibinfo{person}{Yashar Deldjoo}, \bibinfo{person}{Markus
  Schedl}, \bibinfo{person}{Paolo Cremonesi}, {and} \bibinfo{person}{Gabriella
  Pasi}.} \bibinfo{year}{2018}\natexlab{}.
\newblock \showarticletitle{Content-Based Multimedia Recommendation Systems:
  Definition and Application Domains}. In \bibinfo{booktitle}{\emph{Proceedings
  of the 9th Italian Information Retrieval Workshop}}.
\newblock


\bibitem[\protect\citeauthoryear{Deldjoo, Schedl, and Knees}{Deldjoo
  et~al\mbox{.}}{2021c}]%
        {deldjoo2021content}
\bibfield{author}{\bibinfo{person}{Yashar Deldjoo}, \bibinfo{person}{Markus
  Schedl}, {and} \bibinfo{person}{Peter Knees}.}
  \bibinfo{year}{2021}\natexlab{c}.
\newblock \showarticletitle{Content-driven Music Recommendation: Evolution,
  State of the Art, and Challenges}.
\newblock \bibinfo{journal}{\emph{arXiv preprint arXiv:2107.11803}}
  (\bibinfo{year}{2021}).
\newblock


\bibitem[\protect\citeauthoryear{Devlin, Chang, Lee, and Toutanova}{Devlin
  et~al\mbox{.}}{2019}]%
        {DBLP:conf/naacl/DevlinCLT19}
\bibfield{author}{\bibinfo{person}{Jacob Devlin}, \bibinfo{person}{Ming{-}Wei
  Chang}, \bibinfo{person}{Kenton Lee}, {and} \bibinfo{person}{Kristina
  Toutanova}.} \bibinfo{year}{2019}\natexlab{}.
\newblock \showarticletitle{{BERT:} Pre-training of Deep Bidirectional
  Transformers for Language Understanding}. In
  \bibinfo{booktitle}{\emph{Proceedings of the 2019 Conference of the North
  American Chapter of the Association for Computational Linguistics: Human
  Language Technologies, {NAACL-HLT} 2019, Minneapolis, MN, USA, June 2-7,
  2019, Volume 1 (Long and Short Papers)}},
  \bibfield{editor}{\bibinfo{person}{Jill Burstein}, \bibinfo{person}{Christy
  Doran}, {and} \bibinfo{person}{Thamar Solorio}} (Eds.).
  \bibinfo{publisher}{Association for Computational Linguistics},
  \bibinfo{pages}{4171--4186}.
\newblock
\urldef\tempurl%
\url{https://doi.org/10.18653/v1/n19-1423}
\showDOI{\tempurl}


\bibitem[\protect\citeauthoryear{Fern{\'{a}}ndez, Bellog{\'{\i}}n, and
  Cantador}{Fern{\'{a}}ndez et~al\mbox{.}}{2021}]%
        {DBLP:journals/corr/abs-2103-14748}
\bibfield{author}{\bibinfo{person}{Miriam Fern{\'{a}}ndez},
  \bibinfo{person}{Alejandro Bellog{\'{\i}}n}, {and}
  \bibinfo{person}{Iv{\'{a}}n Cantador}.} \bibinfo{year}{2021}\natexlab{}.
\newblock \showarticletitle{Analysing the Effect of Recommendation Algorithms
  on the Amplification of Misinformation}.
\newblock \bibinfo{journal}{\emph{CoRR}}  \bibinfo{volume}{abs/2103.14748}
  (\bibinfo{year}{2021}).
\newblock
\showeprint[arxiv]{2103.14748}
\urldef\tempurl%
\url{https://arxiv.org/abs/2103.14748}
\showURL{%
\tempurl}


\bibitem[\protect\citeauthoryear{Glowacka}{Glowacka}{2019}]%
        {DBLP:journals/ftir/Glowacka19}
\bibfield{author}{\bibinfo{person}{Dorota Glowacka}.}
  \bibinfo{year}{2019}\natexlab{}.
\newblock \showarticletitle{Bandit Algorithms in Information Retrieval}.
\newblock \bibinfo{journal}{\emph{Found. Trends Inf. Retr.}}
  \bibinfo{volume}{13}, \bibinfo{number}{4} (\bibinfo{year}{2019}),
  \bibinfo{pages}{299--424}.
\newblock
\urldef\tempurl%
\url{https://doi.org/10.1561/1500000067}
\showDOI{\tempurl}


\bibitem[\protect\citeauthoryear{Hofmann, Schuth, Bellog{\'{\i}}n, and
  de~Rijke}{Hofmann et~al\mbox{.}}{2014}]%
        {DBLP:conf/ecir/HofmannSBR14}
\bibfield{author}{\bibinfo{person}{Katja Hofmann}, \bibinfo{person}{Anne
  Schuth}, \bibinfo{person}{Alejandro Bellog{\'{\i}}n}, {and}
  \bibinfo{person}{Maarten de Rijke}.} \bibinfo{year}{2014}\natexlab{}.
\newblock \showarticletitle{Effects of Position Bias on Click-Based Recommender
  Evaluation}. In \bibinfo{booktitle}{\emph{Advances in Information Retrieval -
  36th European Conference on {IR} Research, {ECIR} 2014, Amsterdam, The
  Netherlands, April 13-16, 2014. Proceedings}} \emph{(\bibinfo{series}{Lecture
  Notes in Computer Science}, Vol.~\bibinfo{volume}{8416})},
  \bibfield{editor}{\bibinfo{person}{Maarten de~Rijke}, \bibinfo{person}{Tom
  Kenter}, \bibinfo{person}{Arjen~P. de~Vries}, \bibinfo{person}{ChengXiang
  Zhai}, \bibinfo{person}{Franciska de~Jong}, \bibinfo{person}{Kira Radinsky},
  {and} \bibinfo{person}{Katja Hofmann}} (Eds.). \bibinfo{publisher}{Springer},
  \bibinfo{pages}{624--630}.
\newblock
\urldef\tempurl%
\url{https://doi.org/10.1007/978-3-319-06028-6\_67}
\showDOI{\tempurl}


\bibitem[\protect\citeauthoryear{Huang, Oosterhuis, de~Rijke, and van
  Hoof}{Huang et~al\mbox{.}}{2020}]%
        {DBLP:conf/recsys/HuangORH20}
\bibfield{author}{\bibinfo{person}{Jin Huang}, \bibinfo{person}{Harrie
  Oosterhuis}, \bibinfo{person}{Maarten de Rijke}, {and} \bibinfo{person}{Herke
  van Hoof}.} \bibinfo{year}{2020}\natexlab{}.
\newblock \showarticletitle{Keeping Dataset Biases out of the Simulation: {A}
  Debiased Simulator for Reinforcement Learning based Recommender Systems}. In
  \bibinfo{booktitle}{\emph{RecSys 2020: Fourteenth {ACM} Conference on
  Recommender Systems, Virtual Event, Brazil, September 22-26, 2020}},
  \bibfield{editor}{\bibinfo{person}{Rodrygo L.~T. Santos},
  \bibinfo{person}{Leandro~Balby Marinho}, \bibinfo{person}{Elizabeth~M. Daly},
  \bibinfo{person}{Li~Chen}, \bibinfo{person}{Kim Falk}, \bibinfo{person}{Noam
  Koenigstein}, {and} \bibinfo{person}{Edleno~Silva de~Moura}} (Eds.).
  \bibinfo{publisher}{{ACM}}, \bibinfo{pages}{190--199}.
\newblock
\urldef\tempurl%
\url{https://doi.org/10.1145/3383313.3412252}
\showDOI{\tempurl}


\bibitem[\protect\citeauthoryear{Jadidinejad, Macdonald, and Ounis}{Jadidinejad
  et~al\mbox{.}}{2019}]%
        {jadidinejad2019sensitive}
\bibfield{author}{\bibinfo{person}{Amir Jadidinejad}, \bibinfo{person}{Craig
  Macdonald}, {and} \bibinfo{person}{Iadh Ounis}.}
  \bibinfo{year}{2019}\natexlab{}.
\newblock \showarticletitle{How Sensitive is Recommendation Systems' Offline
  Evaluation to Popularity?}. In \bibinfo{booktitle}{\emph{Proceedings of the
  Workshop on Offline Evaluation for Recommender Systems (REVEAL '19),
  co-located with the 13th {ACM} Conference on Recommender Systems (RecSys
  2019)}}.
\newblock


\bibitem[\protect\citeauthoryear{Kluver and Konstan}{Kluver and
  Konstan}{2014}]%
        {DBLP:conf/recsys/KluverK14}
\bibfield{author}{\bibinfo{person}{Daniel Kluver} {and}
  \bibinfo{person}{Joseph~A. Konstan}.} \bibinfo{year}{2014}\natexlab{}.
\newblock \showarticletitle{Evaluating recommender behavior for new users}. In
  \bibinfo{booktitle}{\emph{Eighth {ACM} Conference on Recommender Systems,
  RecSys '14, Foster City, Silicon Valley, CA, {USA} - October 06 - 10, 2014}},
  \bibfield{editor}{\bibinfo{person}{Alfred Kobsa},
  \bibinfo{person}{Michelle~X. Zhou}, \bibinfo{person}{Martin Ester}, {and}
  \bibinfo{person}{Yehuda Koren}} (Eds.). \bibinfo{publisher}{{ACM}},
  \bibinfo{pages}{121--128}.
\newblock
\urldef\tempurl%
\url{https://doi.org/10.1145/2645710.2645742}
\showDOI{\tempurl}


\bibitem[\protect\citeauthoryear{Lesota, Melchiorre, Rekabsaz, Brandl, Kowald,
  Lex, and Schedl}{Lesota et~al\mbox{.}}{2021}]%
        {lesota2021analyzing}
\bibfield{author}{\bibinfo{person}{Oleg Lesota}, \bibinfo{person}{Alessandro
  Melchiorre}, \bibinfo{person}{Navid Rekabsaz}, \bibinfo{person}{Stefan
  Brandl}, \bibinfo{person}{Dominik Kowald}, \bibinfo{person}{Elisabeth Lex},
  {and} \bibinfo{person}{Markus Schedl}.} \bibinfo{year}{2021}\natexlab{}.
\newblock \showarticletitle{Analyzing Item Popularity Bias of Music Recommender
  Systems: Are Different Genders Equally Affected?}. In
  \bibinfo{booktitle}{\emph{Fifteenth ACM Conference on Recommender Systems}}.
  \bibinfo{pages}{601--606}.
\newblock


\bibitem[\protect\citeauthoryear{Lucherini, Sun, Winecoff, and
  Narayanan}{Lucherini et~al\mbox{.}}{2021}]%
        {DBLP:journals/corr/abs-2107-08959}
\bibfield{author}{\bibinfo{person}{Eli Lucherini}, \bibinfo{person}{Matthew
  Sun}, \bibinfo{person}{Amy Winecoff}, {and} \bibinfo{person}{Arvind
  Narayanan}.} \bibinfo{year}{2021}\natexlab{}.
\newblock \showarticletitle{{T-RECS:} {A} Simulation Tool to Study the Societal
  Impact of Recommender Systems}.
\newblock \bibinfo{journal}{\emph{CoRR}}  \bibinfo{volume}{abs/2107.08959}
  (\bibinfo{year}{2021}).
\newblock
\showeprint[arxiv]{2107.08959}
\urldef\tempurl%
\url{https://arxiv.org/abs/2107.08959}
\showURL{%
\tempurl}


\bibitem[\protect\citeauthoryear{Mansoury, Abdollahpouri, Pechenizkiy,
  Mobasher, and Burke}{Mansoury et~al\mbox{.}}{2020}]%
        {DBLP:conf/cikm/MansouryAPMB20}
\bibfield{author}{\bibinfo{person}{Masoud Mansoury}, \bibinfo{person}{Himan
  Abdollahpouri}, \bibinfo{person}{Mykola Pechenizkiy},
  \bibinfo{person}{Bamshad Mobasher}, {and} \bibinfo{person}{Robin Burke}.}
  \bibinfo{year}{2020}\natexlab{}.
\newblock \showarticletitle{Feedback Loop and Bias Amplification in Recommender
  Systems}. In \bibinfo{booktitle}{\emph{{CIKM} '20: The 29th {ACM}
  International Conference on Information and Knowledge Management, Virtual
  Event, Ireland, October 19-23, 2020}},
  \bibfield{editor}{\bibinfo{person}{Mathieu d'Aquin}, \bibinfo{person}{Stefan
  Dietze}, \bibinfo{person}{Claudia Hauff}, \bibinfo{person}{Edward Curry},
  {and} \bibinfo{person}{Philippe Cudr{\'{e}}{-}Mauroux}} (Eds.).
  \bibinfo{publisher}{{ACM}}, \bibinfo{pages}{2145--2148}.
\newblock
\urldef\tempurl%
\url{https://doi.org/10.1145/3340531.3412152}
\showDOI{\tempurl}


\bibitem[\protect\citeauthoryear{Rendle, Freudenthaler, Gantner, and
  Schmidt{-}Thieme}{Rendle et~al\mbox{.}}{2009}]%
        {DBLP:conf/uai/RendleFGS09}
\bibfield{author}{\bibinfo{person}{Steffen Rendle}, \bibinfo{person}{Christoph
  Freudenthaler}, \bibinfo{person}{Zeno Gantner}, {and} \bibinfo{person}{Lars
  Schmidt{-}Thieme}.} \bibinfo{year}{2009}\natexlab{}.
\newblock \showarticletitle{{BPR:} Bayesian Personalized Ranking from Implicit
  Feedback}. In \bibinfo{booktitle}{\emph{{UAI} 2009, Proceedings of the
  Twenty-Fifth Conference on Uncertainty in Artificial Intelligence, Montreal,
  QC, Canada, June 18-21, 2009}}, \bibfield{editor}{\bibinfo{person}{Jeff~A.
  Bilmes} {and} \bibinfo{person}{Andrew~Y. Ng}} (Eds.).
  \bibinfo{publisher}{{AUAI} Press}, \bibinfo{pages}{452--461}.
\newblock
\urldef\tempurl%
\url{https://dslpitt.org/uai/displayArticleDetails.jsp?mmnu=1\&smnu=2\&article\_id=1630\&proceeding\_id=25}
\showURL{%
\tempurl}


\bibitem[\protect\citeauthoryear{Slokom}{Slokom}{2018}]%
        {DBLP:conf/recsys/Slokom18}
\bibfield{author}{\bibinfo{person}{Manel Slokom}.}
  \bibinfo{year}{2018}\natexlab{}.
\newblock \showarticletitle{Comparing recommender systems using synthetic
  data}. In \bibinfo{booktitle}{\emph{Proceedings of the 12th {ACM} Conference
  on Recommender Systems, RecSys 2018, Vancouver, BC, Canada, October 2-7,
  2018}}, \bibfield{editor}{\bibinfo{person}{Sole Pera},
  \bibinfo{person}{Michael~D. Ekstrand}, \bibinfo{person}{Xavier Amatriain},
  {and} \bibinfo{person}{John O'Donovan}} (Eds.). \bibinfo{publisher}{{ACM}},
  \bibinfo{pages}{548--552}.
\newblock
\urldef\tempurl%
\url{https://doi.org/10.1145/3240323.3240325}
\showDOI{\tempurl}


\bibitem[\protect\citeauthoryear{Slokom, Larson, and Hanjalic}{Slokom
  et~al\mbox{.}}{2020}]%
        {slokom2020partially}
\bibfield{author}{\bibinfo{person}{Manel Slokom}, \bibinfo{person}{Martha
  Larson}, {and} \bibinfo{person}{Alan Hanjalic}.}
  \bibinfo{year}{2020}\natexlab{}.
\newblock \showarticletitle{Partially Synthetic Data for Recommender Systems:
  Prediction Performance and Preference Hiding}.
\newblock \bibinfo{journal}{\emph{arXiv preprint arXiv:2008.03797}}
  (\bibinfo{year}{2020}).
\newblock


\bibitem[\protect\citeauthoryear{Valcarce, Bellog{\'{\i}}n, Parapar, and
  Castells}{Valcarce et~al\mbox{.}}{2018}]%
        {DBLP:conf/recsys/ValcarceBPC18}
\bibfield{author}{\bibinfo{person}{Daniel Valcarce}, \bibinfo{person}{Alejandro
  Bellog{\'{\i}}n}, \bibinfo{person}{Javier Parapar}, {and}
  \bibinfo{person}{Pablo Castells}.} \bibinfo{year}{2018}\natexlab{}.
\newblock \showarticletitle{On the robustness and discriminative power of
  information retrieval metrics for top-N recommendation}. In
  \bibinfo{booktitle}{\emph{Proceedings of the 12th {ACM} Conference on
  Recommender Systems, RecSys 2018, Vancouver, BC, Canada, October 2-7, 2018}},
  \bibfield{editor}{\bibinfo{person}{Sole Pera}, \bibinfo{person}{Michael~D.
  Ekstrand}, \bibinfo{person}{Xavier Amatriain}, {and} \bibinfo{person}{John
  O'Donovan}} (Eds.). \bibinfo{publisher}{{ACM}}, \bibinfo{pages}{260--268}.
\newblock
\urldef\tempurl%
\url{https://doi.org/10.1145/3240323.3240347}
\showDOI{\tempurl}


\end{thebibliography}

\end{document}